\documentstyle [prl,aps,preprint]{revtex}

\begin{document}
\draft
\title{Extending the quantal adiabatic theorem: Geometry of 
noncyclic motion\footnote{Accepted in American Journal of Physics}}
\author{Gonzalo Garc\'{\i}a de 
Polavieja$^{1,}$\footnote{E-mail: 
gonzalo@rydberg.thchem.ox.ac.uk} 
and Erik Sj\"{o}qvist$^{2,}$\footnote{E-mail: 
erik.sjoqvist@philosophy.oxford.ac.uk}}
\address{$^{1}$ Physical and Theoretical Chemistry 
Laboratory, South Parks Road, Oxford OX1 3QZ, UK \\
$^{2}$Sub-Faculty of Philosophy, Oxford University, 
10 Merton Street, Oxford OX1 4JJ, UK}
\maketitle
\begin{abstract}
We show that a noncyclic phase of geometric origin has 
to be included in the approximate adiabatic wave function. 
The adiabatic noncyclic geometric phase for systems 
exhibiting a conical intersection as well as for an 
Aharonov-Bohm situation is worked out in detail. A 
spin$-\frac{1}{2}$ experiment to measure the adiabatic 
noncyclic geometric phase is discussed. We also analyze 
some misconceptions in the literature and textbooks 
concerning noncyclic geometric phases. 
\end{abstract}
\section{Introduction}
Imagine we slowly modify the state of a quantum system with 
some external parameters. A paradigmatic example could be 
a particle with spin in a slowly rotating magnetic field. 
The adiabatic change could be very large, but it takes 
place over a long time such that the transition between 
different energy levels is negligible. What is the 
approximate form of the adiabatic wave function? This 
question seemed to be settled since the first treatment 
of Born and Fock \cite{born} in quantum mechanics that 
extended previous work of Ehrenfest \cite{ehrenfest} 
in classical mechanics and the old quantum theory. 
Surprisingly, a new insight into the adiabatic theorem 
had to wait 55 years to Berry's \cite{berry1} 1984 paper. 
Treatments of the adiabatic approximation before Berry's 
analysis correctly demonstrated that for a slowly varying 
Hamiltonian $\hat{H}(t)$ the solution of the Schr\"{o}dinger 
equation 
\begin{equation}
i\hbar | \dot{\Psi} \rangle = \hat{H} |\Psi \rangle 
\label{eq:se}
\end{equation}
for a state vector initially in $|\Psi ;0 \rangle = |n;0\rangle$ 
is within the adiabatic approximation of the form 
\cite{dbohm,messiah,schiff}
\begin{equation}
|\Psi ;t \rangle = 
\exp \left( i\alpha_{n} (t) -\frac{i}{\hbar} \int_{0}^{t}dt'  
E_{n}(t') \right) |n;t\rangle ,  
\label{eq:wrongsolution}
\end{equation}
where $|n;t\rangle$ is an instantaneous eigenstate of 
$\hat{H} (t)$
\begin{equation}
\hat{H} (t) |n;t\rangle = E_{n} (t) |n;t\rangle .
\end{equation}
However, these treatments failed to realize the relevance 
of the phase $\alpha_{n}$ as they were more interested in 
obtaining the conditions of validity of (\ref{eq:wrongsolution}) 
namely that no transition between levels will occur if 
\begin{equation}
\hbar \frac{|\langle n;t| \dot{\hat{H}} |m;t\rangle |}
{|E_{n}(t)-E_{m}(t)|^{2}} \ll 1. 
\label{eq:advalidity}
\end{equation}

Berry showed that if the system is transported around a 
closed circuit $C_{0}$ in parameter space in time $T$ by 
changing the parameters ${\bf R}$ in the Hamiltonian 
$\hat{H}({\bf R})$, the final wave function is of the form 
\begin{equation}
|\Psi ;T \rangle = \exp \left( i\gamma _{n} [C_{0}]
-\frac{i}{\hbar} \int_{0}^{T} dt' E_{n}({\bf R}(t')) 
\right) |\Psi ;0 \rangle  
\label{eq:correctsolution}
\end{equation}
with the geometrical phase change
\begin{equation}
\gamma_{n} [C_{0}] =i\oint_{C_{0}}d{\bf R}\cdot \langle
n;{\bf R} | \nabla_{{\bf R}} |n;{\bf R} \rangle , 
\label{eq:cyclicgp}
\end{equation}
where the instantaneous nondegenerate energy eigenstate 
$|n;{\bf R} \rangle$ is singlevalued in a parameter space 
domain that includes the closed circuit $C_{0}$. From 
(\ref{eq:cyclicgp}) is clear that $\gamma_{n} [C_{0}]$ 
is independent of how the circuit is traversed provided the
traversal is slow enough for the adiabatic approximation to 
hold. Using Stokes' theorem the geometric phase in 
(\ref{eq:cyclicgp}) was written by Berry as 
\begin{equation}
\gamma_{n} [C_{0}] = 
-\int_{S_{0}}d{\bf S}\cdot V_{n}({\bf R}) ,  
\label{eq:2-formgp}
\end{equation}
with 
\begin{equation}
V_{n}({\bf R}) = {\mbox{Im}} \sum_{m\neq n}
\frac{\langle n;{\bf R} |(\nabla_{{\bf R}}
\hat{H} ({\bf R}))| m;{\bf R} \rangle \times
\langle m;{\bf R} |(\nabla_{{\bf R}}
\hat{H} ({\bf R}))| n;{\bf R} \rangle}
{( E_{n}({\bf R})-E_{m}({\bf R}))^{2}} .
\label{eq:2-form}
\end{equation}
Clearly expression (\ref{eq:2-formgp}) for the geometric 
phase $\gamma_{n} [C_{0}] $ is independent of the phase 
choice of $|n;{\bf R}\rangle$ and therefore there is 
no need to choose a singlevalued  $|n;{\bf R} \rangle$ as 
in relation (\ref{eq:cyclicgp}). This property and the 
independence of how the path $C_{0}$ is traversed in 
parameter space are the geometric properties of 
$\gamma_{n} [C_{0}]$, known as Berry's geometric phase. 
This is precisely the relevance of Berry's result, that
$\gamma_{n} [C_{0}]$ is of geometric origin. 

Berry's discovery of the geometric phase for cyclic 
adiabatic evolution has met several generalizations. 
Aharonov and Anandan \cite{aharonov1} generalized 
Berry's phase to nonadiabatic cyclic motion. Samuel 
and Bhandari \cite{samuel} obtained the geometric 
phase for noncyclic cases. Further generalizations
have been the purely kinematic approaches of Aitchison 
and Wanelik \cite{aitchison1} and Mukunda and Simon 
\cite{mukunda}. These works are mostly concerned with 
the geometric phase of exact solutions of the 
Schr\"{o}dinger equation and have introduced a 
geometrical understanding of quantum theory.

In this paper we are concerned with approximate solutions 
for quantum adiabatic motion. Treatments of the adiabatic 
change in quantum mechanics after Berry's result for cyclic 
adiabatic evolution stress the necessity of cyclicity 
\cite{aitchison2,jackiw} and even in some cases explicitly 
state that the phase additional to the dynamical phase 
$-\int_{0}^{T} dt' E_{n} (t')/\hbar$ can be eliminated 
\cite{niemi,abohm} in the noncyclic adiabatic case but not in
the cyclic one. Is cyclic adiabatic motion so special? Is it 
necessary for the appearance of the adiabatic geometric phase? 
It is the purpose of this paper to show that the geometric phase 
also appears in noncyclic adiabatic motion and that there is 
nothing special in the cyclic case for the existence of the  
geometric phase.

We have organized this paper as follows. In Sec. {\bf II} 
a derivation of the geometric phase for adiabatic noncyclic  
change is given. We show that this phase cannot be canceled 
and that it is geometric. It coincides with Berry's phase for 
cyclic evolutions. Secs. {\bf III}$-${\bf V} discuss examples 
of noncyclic geometric phases as well as implications for 
experiments.

\section{Noncyclic adiabatic change}
In this section we show that the correct treatment of noncyclic 
adiabatic change also needs the inclusion of a phase of 
geometric origin. This geometric contribution reduces to 
Berry's phase for cyclic evolution. We assume the same 
conditions as in Berry's derivation of discrete nondegenerate 
spectrum and Hermitian Hamiltonian. In the noncyclic case, 
${\bf R}$ traces out an open path $C$ in parameter space.  
Under adiabatic evolution the solution of the Schr\"{o}dinger 
equation (\ref{eq:se}) for a state initially in 
$|\Psi ;0 \rangle = |n;{\bf R}(0) \rangle$ is approximately 
\begin{equation}
|\Psi ;t \rangle = 
\exp \left( i\alpha_{n}(t)-\frac{i}{\hbar} 
\int_{0}^{t} dt' E_{n}({\bf R}(t')) \right) 
|n;{\bf R}(t) \rangle ,   
\label{eq:adsolution}
\end{equation}
where $|n;{\bf R}(t)\rangle$ is the nth instantaneous energy 
eigenstate. Substitution of (\ref{eq:adsolution}) in 
Schr\"{o}dinger's equation (\ref{eq:se}) gives the  
phase $\alpha_{n} (t)$ as 
\begin{equation}
\alpha_{n}(t) - \alpha_{n}(0) = 
i\int_{0}^{t}dt' \, \dot{{\bf R}} (t') \cdot
\langle n;{\bf R}(t') |\nabla_{{\bf R}}| n;{\bf R}(t') \rangle , 
\label{eq:dynamicalphase}
\end{equation}
which is real as the basis $|n;{\bf R}(t)\rangle$ is
normalized. Multiplying (\ref{eq:adsolution}) by 
$\langle \Psi ;0 | = \langle n;{\bf R}(0)|$ yields 
\begin{equation}
\arg \langle \Psi ;0 | \Psi ;t \rangle =
\gamma_{n} [C] - \frac{1}{\hbar}\int_{0}^{t} dt' E_{n}({\bf R}(t')) 
\label{eq:generaltp}
\end{equation}
with $\gamma_{n} [C]$ the adiabatic noncyclic 
phase of the form 
\begin{equation}
\gamma_{n} [C] = 
\arg \langle n; {\bf R}(0) | n;{\bf R}(t)\rangle  
+ i\int_{0}^{t}dt' \, \dot{{\bf R}} (t') \cdot
\langle n;{\bf R}(t') |\nabla_{{\bf R}}| n;{\bf R}(t') \rangle ,   
\label{eq:generalgp}
\end{equation}
which is defined when the eigenvectors 
$|n;{\bf R}(0) \rangle$ and $|n;{\bf R}(t) \rangle$  
are nonorthogonal. 

Expression (\ref{eq:generalgp}) is the general expression for 
the adiabatic noncyclic geometric phase that only depends on the
open path $C$ in parameter space. Its geometric nature is 
explained in the following. The adiabatic noncyclic geometric 
phase in (\ref{eq:generalgp}) is independent of the phase 
choice of $|n;{\bf R}(t)\rangle$ as it is invariant 
under the global phase transformation $|n;{\bf R}(t)\rangle 
\longrightarrow \exp \left( i\lambda ({\bf R}(t)) \right)
|n;{\bf R}(t)\rangle$. Therefore $\gamma_{n} [C]$
cannot be eliminated by a global phase transformation and has 
to be included in any calculation for adiabatic evolution (the 
invariance of the geometric phase for other unitary transformations
has been discussed in the literature 
\cite{aharonov1,berry2,giavarini,sjoqvist,garcia}). The 
adiabatic noncyclic geometric phase is also independent of 
how the path is traversed as any differentiable monotonic 
transformation $t\longrightarrow \nu (t)$ leaves 
(\ref{eq:generalgp}) invariant. An alternative to the expression 
for the geometric phase in (\ref{eq:generalgp}) is given by 
the approach that uses a geodesic closure of the path 
\cite{samuel}. The disadvantages of this noncyclic approach 
for the adiabatic theorem are explained in the Appendix. 

Using (\ref{eq:adsolution}) and (\ref{eq:generaltp}) 
for a cyclic evolution we can write the state vector 
at time $T$ as in (\ref{eq:correctsolution})
with $\gamma_{n} [C_{0}]$ the adiabatic geometric phase for 
a cyclic path $C_{0}$ writing (\ref{eq:generalgp}) for the 
cyclic time $T$. Obviously, as in the noncyclic case, the 
adiabatic cyclic geometric phase $\gamma_{n} [C_{0}]$ is 
globally phase invariant and independent on how the circuit 
$C_{0}$ is traversed. Berry's result is immediately obtained by 
making a global phase transformation to a singlevalued basis, 
for which $|n;{\bf R}(T) \rangle = |n;{\bf R}(0)\rangle$, and 
it can clearly be seen that relation (\ref{eq:generalgp}) 
reduces to Berry's expression (\ref{eq:cyclicgp}).

We would like to highlight briefly why the adiabatic 
noncyclic geometric phase might have been overlooked 
in the past. First, Berry's expressions for the geometric 
phase in (\ref{eq:cyclicgp}) and (\ref{eq:2-formgp}) are 
clearly independent of how the path in parameter space is 
traversed and are also invariant under global phase 
transformations. These properties that define the geometric 
phase are less obvious at first sight in the noncyclic 
case as discussed above. Secondly, the first treatments 
and reviews of the geometric phase stressed the theoretical 
and experimental importance of cyclicity 
\cite{berry1,aharonov1,aitchison2,jackiw,simon}. Thirdly, 
several incorrect statements have been given in the literature, 
which might have caused confusion about the significance
of the adiabatic noncyclic geometric phase. For example, 
it has been stated that there is no adiabatic noncyclic 
geometric phase \cite{niemi,abohm} or that it is given by 
(\ref{eq:dynamicalphase}) instead of (\ref{eq:generalgp}) 
\cite{wu}. Moreover, the nonadiabatic noncyclic treatment 
of Samuel and Bhandari \cite{samuel} was very formal and no 
explicit physical examples were given in their article. The 
kinematic approach of Aitchison and Wanelik \cite{aitchison1} 
made noncyclic geometric phases more accessible but again no 
explicit examples were given. In the kinematic study of 
Mukunda and Simon \cite{mukunda} examples in the field of 
optics were given and the adiabatic noncyclic geometric 
phase in an optical set-up has been correctly calculated 
by Christian and Shimony \cite{christian} using the 
kinematic approach.

We have shown in this section that a phase of geometric 
origin has to be included in the Schr\"{o}dinger solution 
for an adiabatic noncyclic evolution. This adiabatic noncyclic 
geometric phase cannot be eliminated by a global phase 
transformation as it is independent of the phase choice of 
the basis and reduces to Berry's adiabatic cyclic phase for 
cyclic times.

\section{Noncyclic perspective of the sign change problem}
It is well-known in molecular physics since the work by 
Longuet-Higgins and coworkers \cite{longuet,herzberg} 
that adiabatically transporting the electronic state 
around a conical intersection in nuclear configuration 
space changes the sign of the electronic wave function. 
Therefore the nuclear wave function must also change sign 
to make the total product wave function singlevalued. 
This effect has been observed in the vibrational 
spectrum of Na$_{3}$ \cite{delacretaz}. We here analyze 
the sign change problem from a noncyclic perspective. 
Specifically, consider a two-dimensional subspace of 
nuclear configuration space in which two electronic 
states conically intersect at the origin. In a 
neighborhood of the degenerate point the Hamiltonian 
can, up to an additive multiple of the unit matrix, 
be written as a real symmetric $2\times 2$ matrix 
\cite{mead1,mead2}  
\begin{equation}
\hat{H} = R (\sin \Phi \, \hat{\sigma}_{x} + 
\cos \Phi \, \hat{\sigma}_{z}) ,
\label{eq:realhamiltonian}
\end{equation}
where $\hat{\sigma}_{x}$ and $\hat{\sigma}_{z}$ are the 
$x-$ and $z-$components of the usual Pauli matrices 
\begin{equation}
\hat{\sigma}_{x} = 
\left( \begin{array}{cc} 
0 & \, 1 \\ 1 & \, 0 
\end{array} \right) \! , \, \, 
\hat{\sigma}_{y} = 
\left( \begin{array}{cc} 
0 & -i \\ i & 0 
\end{array} \right) \! , \, \, 
\hat{\sigma}_{z} = 
\left( \begin{array}{cc} 
1 & 0 \\ 0 & -1 
\end{array} \right)  
\label{eq:pauli}
\end{equation}
and the parameters $(R,\Phi )$, which are here treated 
as classical, are nuclear polar coordinates. It can 
be checked that  
\begin{eqnarray}
|+;\Phi \rangle & = & 
\left( \begin{array}{cc} \cos (\Phi /2) \\ 
\sin (\Phi /2) \end{array} \right) 
\nonumber \\ 
|-;\Phi \rangle & = & 
\left( \begin{array}{cc} -\sin (\Phi /2) \\ 
\cos (\Phi /2) \end{array} \right) 
\label{eq:realeigenstates} 
\end{eqnarray}
are eigenvectors of $\hat{H}$ with the corresponding 
energy eigenvalues
\begin{equation}
E_{\pm} (R) = \pm R .  
\label{eq:conical}
\end{equation}
Clearly the energies $E_{\pm} (R)$ conically intersect  
at $R=0$, as shown in Fig. 1. 

Now suppose the nuclear motion is such that $R$ is 
nonzero and constant but $\Phi$ varies slowly with 
time, i.e. $|\dot{\Phi}|\ll 4R/\hbar$ from 
(\ref{eq:advalidity}). Taking the initial state to be  
\begin{equation}
|\Psi ;0 \rangle = |+;\Phi (0) \rangle , 
\end{equation}
where $|+\rangle$ is given by (\ref{eq:realeigenstates}), 
it follows from (\ref{eq:adsolution}), 
(\ref{eq:dynamicalphase}) and (\ref{eq:conical}) that  
\begin{equation}
|\Psi ;t \rangle = 
\exp \left( -\frac{i}{\hbar} Rt \right) 
|+;\Phi (t)\rangle  
\label{eq:sesolution}
\end{equation}
is a solution of the time dependent Schr\"{o}dinger 
equation in the adiabatic approximation. The geometric 
phase of the solution $|\Psi ;t \rangle$ accumulated 
during the interval $[0,t]$ can be calculated from 
(\ref{eq:generalgp}) using 
(\ref{eq:realeigenstates}) 
\begin{eqnarray}
\gamma_{+} [C] & = &  
\arg \langle +;\Phi (0)|+;\Phi (t) \rangle + 0 = 
\arg \cos ((\Phi (t)-\Phi (0))/2)  
\nonumber \\ 
 & = & \left\{ \begin{array}{ll} 
0 & {\mbox{if}} \, \, 
\Phi (0)\leq \Phi (t) < \Phi (0)+\pi \\
{\mbox{undefined}} & {\mbox{if}} \, \, 
\Phi (t) = \Phi (0)+\pi \\ 
\pi & {\mbox{if}} \, \, 
\Phi (0)+\pi < \Phi (t) \leq \Phi (0)+2\pi 
\end{array} \right. .
\label{eq:realgp}
\end{eqnarray}
This shows that the adiabatic noncyclic geometric phase 
is nonzero. It also shows that the sign change of the 
cyclic electronic wave function, first noticed by 
Longuet-Higgins and coworkers \cite{longuet,herzberg},  
is due to the sign change of the noncyclic geometric 
phase factor at $\Phi (t) = \Phi (0) + \pi$. 

From the invariance under global phase transformations 
and independence of how the path is traversed, discussed 
below (\ref{eq:generalgp}) in Sec. {\bf II}, it is clear 
that the sign change must also follow if we choose an 
eigenvector which can be complex and/or nondifferentiable 
at some isolated points along $C$. It is instructive to 
demonstrate this explicitly. We introduce, by global phase 
transformations of $|+;\Phi (t)\rangle$, two new vectors 
which are labeled as $|+;\Phi (t) \rangle'$ and 
$|+;\Phi (t) \rangle''$. Take first the complex vector  
\begin{equation}
|+;\Phi (t)\rangle' = 
\exp \left( i\Phi (t)/2 \right) 
|+;\Phi (t)\rangle ,  
\end{equation}
which is differentiable and singlevalued in $\Phi$. We 
then have
\begin{equation}
'\langle +;\Phi (t)|
\frac{1}{R}\frac{\partial}{\partial \Phi}
|+;\Phi (t) \rangle' = \frac{i}{2R}  
\label{eq:dynamicalcomplex}
\end{equation}
and
\begin{equation}
'\langle +;\Phi (0) |+;\Phi (t) \rangle' =
\exp \left( i(\Phi (t)-\Phi (0))/2 \right) 
\langle +;\Phi (0) |+;\Phi (t) \rangle . 
\label{eq:totalcomplex}
\end{equation}
Inserting (\ref{eq:dynamicalcomplex}) and 
(\ref{eq:totalcomplex}) into (\ref{eq:generalgp}) we obtain the 
geometric phase 
\begin{eqnarray}
\gamma'_{+} [C] & = & \frac{1}{2} (\Phi (t)-\Phi (0)) +  
\arg \langle +;\Phi (0)|+;\Phi (t) \rangle - 
\frac{1}{2} \int_{\Phi (0)}^{\Phi (t)} d\Phi 
\nonumber \\ 
 & = & \arg \langle +;\Phi (0)|+;\Phi (t) \rangle  = 
\gamma_{+} [C] . 
\end{eqnarray}

Next let us consider the real vector  
\begin{equation}
|+;\Phi (t)\rangle'' = 
\exp \left( -i \arg 
\langle +;\Phi (0)|+;\Phi (t) \rangle \right)  
|+;\Phi (t)\rangle , 
\label{eq:singlevalued}
\end{equation}
which is singlevalued but 
not differentiable everywhere as it has a finite jump 
at $\Phi (t) = \Phi (0)+\pi$. We now have 
\begin{equation}
\arg (''\langle +;\Phi (0)|+;\Phi (t) \rangle'' ) = 
\left\{ \begin{array}{ll} 
0 & {\mbox{if}} \, \, \Phi (t) \neq \Phi (0)+\pi \\
{\mbox{undefined}} & {\mbox{if}} \, \, \Phi (t) = \Phi (0)+\pi 
\end{array} \right.  
\label{eq:totalnondiff}
\end{equation}
and at $\Phi (t) \neq \Phi (0)+\pi$ 
\begin{equation}
''\langle +;\Phi (t)|
\frac{1}{R}\frac{\partial}{\partial \Phi}
| +;\Phi (t) \rangle'' = 
-\frac{i}{R} \frac{\partial}{\partial \Phi} \arg 
\langle +;\Phi (0)|+;\Phi (t) \rangle . 
\label{eq:dynamicalnondiff}
\end{equation}
Inserting (\ref{eq:totalnondiff}) and (\ref{eq:dynamicalnondiff}) 
into (\ref{eq:generalgp}) yields 
\begin{equation}
\gamma''_{+} [C] = 0 + \int_{\Phi (0)}^{\Phi (t)} 
\frac{\partial}{\partial \Phi} \arg \langle 
+;\Phi (0)|+;\Phi (t') \rangle d\Phi = 
\gamma_{+} [C] .
\end{equation}

We have seen in this section that a simple case of degeneracy 
provides a clear example of the adiabatic noncyclic geometric 
phase and its properties. It has also been shown that the 
adiabatic noncyclic geometric phase provides a detailed 
characterization of the sign change problem. In the particular 
case chosen here we have shown that the sign change is due to 
an adiabatic noncyclic geometric phase of value $\pi$ after  
$\Phi (0) + \pi$.

\section{Relationship between the Aharonov-Bohm effect and the 
noncyclic geometric phase}
Quantum theory has been given several surprises after it
was completed in the early 30s. One of them is precisely 
the geometric phase. Another surprising result was discovered 
by Aharonov and Bohm in 1959 \cite{aharonov2}. This section 
concerns the relation between the Aharonov-Bohm effect and 
the noncyclic geometric phase. 

Consider a particle with charge 
$q$ confined in an impenetrable box transported along a circuit 
$C$ around a magnetic flux line  as shown in Fig. 2. The 
relative coordinate between the box and a point on the flux 
line is ${\bf R}$. An asymmetric shape of the box has been 
chosen in order to avoid energy degeneracies. The size of the
box in Fig. 2 is such that the overlap between the initial wave
function and the wave function at any later time $t$ is 
nonvanishing which guarantees that the noncyclic geometric 
phase is well defined except at isolated ${\bf R}-$values where 
these wave functions might be orthogonal. 

Suppose $\psi_{n} ({\bf r};{\bf R} (t))$ 
is the real-valued energy eigenfunction in absence of 
magnetic flux associated with the ${\bf R}-$independent 
energy $E_{n}$. Turning on the flux the eigenfunction  
becomes  
\begin{equation}
\varphi_{n} ({\bf r};{\bf R} (t)) = 
\exp \left( i\frac{q}{\hbar c} 
\int_{{\bf R}(t)}^{{\bf r}} d{\bf r}' \cdot {\bf A} ({\bf r}') 
\right) \psi_{n} ({\bf r};{\bf R}(t))   
\label{eq:abeigenfunctions}
\end{equation}
and the energy is still $E_{n}$. In cylindrical 
coordinates $(r,\theta ,z)$ the vector potential reads 
\begin{equation}
{\bf A} = \frac{\Lambda}{2\pi r} {\bf e}_{\theta} , 
\label{eq:abpotential}
\end{equation}
where $\Lambda$ is the flux and ${\bf e}_{\theta}$ is 
the unit vector in the $\theta-$direction. Introducing the 
dimensionless parameter $\eta = q\Lambda / (2\pi \hbar c)$, 
the eigenfunction (\ref{eq:abeigenfunctions}) becomes 
\begin{equation}
\varphi_{n} ({\bf r};{\bf R}(t)) = 
\exp \left( i\eta \int_{{\bf R}(t)}^{{\bf r}} d\theta'  
\right) \psi_{n} ({\bf r};{\bf R}(t)) ,   
\label{eq:abeigfunc}
\end{equation}
which has to be singlevalued in ${\bf r}$ for every ${\bf R} (t) = 
(R\cos \Theta (t),R\sin \Theta (t),Z)$. In the following we analyze
the noncyclic geometric phase by dividing the transport of the box 
into three different cases which correspond to three different 
intervals of $\Theta (t)$, as shown in Fig. 2. For notational 
convenience we write the eigenfunctions and the integrals only 
in terms of $\theta$ and $\Theta$, and restrict $\theta$ to the 
interval $[0,2\pi [$.

{\it Case (a).} ($0 \leq \Theta (t) \leq 2\pi - \Delta \theta$) 
The eigenfunction (\ref{eq:abeigfunc}) for this case can be 
written as 
\begin{equation}
\varphi_{n} (\theta ;\Theta (t)) = \left\{  
\begin{array}{ll} 
0 & {\mbox{if}} \, \, 0 \leq \theta \leq \Theta (t) \\ 
\exp \left( i\eta (\theta - \Theta (t)) \right) 
\psi_{n} (\theta -\Theta (t)) &  
{\mbox{if}} \, \, \Theta (t) \leq \theta \leq \Theta (t) + \Delta \theta \\ 
0 & {\mbox{if}} \, \, \Theta (t) + \Delta \theta \leq \theta < 2\pi , 
\end{array}
\right. 
\end{equation}
where $\Delta \theta$ is the angular length of the box. 
First we calculate 
\begin{equation}
\arg \langle \varphi_{n} ; 0 | \varphi_{n} ; \Theta (t) \rangle =   
\arg \int_{\Theta (t)}^{\Delta \theta} d\theta  
\exp \left( -i \eta \Theta (t) \right) 
\psi_{n} (\theta ) \psi_{n} (\theta -\Theta (t)) = 
- \eta \Theta (t) .  
\label{eq:abtotalphase_a}  
\end{equation}
Here we have used that $\psi_{n}$ is real-valued and have chosen  
$\Theta (0) = 0$. Furthermore we have 
\begin{eqnarray}
\langle \varphi_{n} ;\Theta (t) | \nabla_{{\bf R}} | 
\varphi_{n} ;\Theta (t) \rangle & = & \int_{0}^{2\pi} d\theta  
\left( -i \frac{q}{\hbar} {\bf A} ({\bf R}(t)) 
\psi_{n}^{2} (\theta -\Theta (t)) \right.  
\nonumber \\ 
 & & + \psi_{n} (\theta -\Theta (t)) \nabla_{{\bf R}}  
\psi_{n} (\theta -\Theta (t)) \left) = 
-i\frac{\eta}{R} {\bf e}_{\Theta} + 0 , \right.  
\label{eq:abdynamicalphase}
\end{eqnarray}
where the second term on the r.h.s. vanishes as $\psi_{n}$ is 
normalized and real-valued. Inserting (\ref{eq:abtotalphase_a}) 
and (\ref{eq:abdynamicalphase}) into (\ref{eq:generalgp}) we 
obtain the noncyclic geometric phase for case (a) as  
\begin{equation}
\gamma_{n} [C] = -\eta \Theta (t) + 
\int_{0}^{\Theta (t)} Rd\Theta \frac{\eta}{R} = 0 .
\end{equation}

{\it Case (b).} ($2\pi - \Delta \theta \leq \Theta (t) \leq \Delta \theta$)
The eigenfunction for this case that is singlevalued in ${\bf r}$   
is of the form 
\begin{equation}
\varphi_{n} (\theta ;\Theta (t)) = \left\{ 
\begin{array}{ll} 
\exp \left( i\eta 
(\theta - \Theta (t) + 2\pi ) \right) 
\psi_{n} (\theta -\Theta (t) + 2\pi ) & 
{\mbox{if}} \, \, 0 \leq \theta \leq \Theta (t) + \Delta \theta -2\pi \\ 
0 & {\mbox{if}} \, \, \Theta (t) + \Delta \theta -2\pi \leq 
\theta \leq \Theta (t) \\ 
\exp \left( i \eta  (\theta - \Theta (t) ) \right) 
\psi_{n} (\theta -\Theta (t)) & 
{\mbox{if}} \, \, \Theta (t) \leq \theta < 2\pi . 
\label{eq:abeigfunc_b} 
\end{array}
\right. 
\end{equation}
Using this eigenfunction we find 
\begin{eqnarray}
\arg \langle \varphi_{n} ; 0 |
\varphi_{n} ; \Theta (t) \rangle & = & 
- \eta \Theta (t) + \arg 
\left( \exp \left( i 2 \pi \eta \right) 
\int_{0}^{\Theta (t) + \Delta \theta - 2\pi} d\theta \, 
\psi_{n} (\theta ) \psi_{n} (\theta - \Theta (t) + 2\pi ) 
\right. \nonumber \\
 & & + \left. \int_{\Theta (t)}^{\Delta \theta} d\theta \,  
\psi_{n} (\theta ) \psi_{n} (\theta - \Theta (t)) 
\right)   
\label{eq:abtotalphase_b} 
\end{eqnarray}
and the second term in (\ref{eq:generalgp}) coincides with 
(\ref{eq:abdynamicalphase}) of case (a). The noncyclic geometric
phase is then
\begin{eqnarray}
\gamma [C] & = &  
\arg \left( \exp \left( i 2\pi \eta \right) 
\int_{0}^{\Theta (t) + \Delta \theta - 2\pi} d\theta \, 
\psi_{n} (\theta ) \psi_{n} (\theta - \Theta (t) + 2\pi )  
\right. \nonumber \\  
& & + \left. \int_{\Theta (t)}^{\Delta \theta} d\theta \, 
\psi_{n} (\theta ) \psi_{n} (\theta - \Theta (t)) 
\right) .  
\label{eq:abgp_b} 
\end{eqnarray}

{\it Case (c).} ($\Theta (t) \geq \Delta \theta$)
The eigenfunction for this case is again given by 
(\ref{eq:abeigfunc_b}). In fact we can obtain its corresponding 
noncyclic geometric phase by taking the limit $\Theta \longrightarrow
\Delta \theta$ in (\ref{eq:abgp_b}) yielding 
\begin{equation}
\gamma [C] = 2 \pi \eta . 
\label{eq:abgp_c} 
\end{equation}
Hence the noncyclic geometric phase coincides with the Aharonov-Bohm 
phase for this $\Theta -$interval. 

A special case of (c) is the cyclic transport discussed by Berry 
\cite{berry1}, and which can also be found in a later addition to 
Sakurai's book \cite{sakurai} and in Refs. 
\cite{kobe,nagoshi,morandi,dittrich,holstein}. Note however that the 
noncyclic transport shows a richer relation between the geometric 
phase and the Aharonov-Bohm effect. That is, the noncyclic geometric 
phase interpolates smoothly between $0$ (case (a)) and the Aharonov-Bohm 
value $2\pi \eta$ (case (c)) as shown by (\ref{eq:abgp_b}). It is in 
principle possible to experimentally verify this from the interference 
in the overlapping regions between a box transported along $C$ and another 
one not transported as they both would have the same dynamical phase.

\section{Noncyclic neutron polarization experiment}
The geometric phase for adiabatic cyclic 
evolutions has been observed for various physical 
systems such as neutrons \cite{bitter} and photons 
\cite{kwiat}. Is it also possible to experimentally 
verify the noncyclic geometric phase? A claim that 
this has been done appeared in a paper by Weinfurter 
and Badurek \cite{weinfurter}, but it has recently been 
pointed out \cite{wagh1} that the phase they observed is 
not of geometric origin. Correct theoretical proposals of 
how to observe the noncyclic geometric phase have been 
put forward \cite{christian,wagh1,wagh2}. Here we give 
an elementary demonstration of how the adiabatic 
noncyclic geometric phase can be measured. 

Consider a neutron beam subject to a homogeneous magnetic 
field with constant magnitude $B$. Its direction, given by 
the unit vector ${\bf e}$, varies slowly. Denoting the 
magnetic moment of the neutrons by $\mu$, the spin-Hamiltonian 
reads    
\begin{equation}
\hat{H} = -\mu B {\bf e} (t) \cdot {\underline{\hat{\sigma}}} , 
\label{eq:spinhamiltonian}
\end{equation}
where ${\underline{\hat{\sigma}}} = 
(\hat{\sigma}_{x} ,\hat{\sigma}_{y} ,\hat{\sigma}_{z} )$ 
are the Pauli matrices (\ref{eq:pauli}). Writing the unit 
vector ${\bf e} (t)$ in spherical coordinates  
\begin{equation}
{\bf e} (t) = (\sin \Theta (t) \cos \Phi (t) ,
\sin \Theta (t) \sin \Phi (t) ,\cos \Theta (t))
\end{equation}
the instantaneous eigenstates of 
(\ref{eq:spinhamiltonian}) can be expressed as 
\begin{eqnarray}
|+;{\bf e} (t) \rangle & = &  
\exp (-i\Phi (t)/2) \cos (\Theta (t)/2) 
|+;{\bf e}_{z} \rangle + 
\exp (i\Phi (t)/2) \sin (\Theta (t)/2) 
|-;{\bf e}_{z} \rangle 
\nonumber \\
|-;{\bf e} (t) \rangle & = &  
-\exp (-i\Phi (t)/2) \sin (\Theta (t)/2)
|+;{\bf e}_{z} \rangle + 
\exp (i\Phi (t)/2) \cos (\Theta (t)/2) 
|-;{\bf e}_{z} \rangle , 
\label{eq:complexeigenstates}
\end{eqnarray}
with the corresponding energies 
\begin{equation}
E_{\pm} (B) = \mp \mu B \equiv \mp \hbar \omega_{B} .
\end{equation}
We see that the energies conically intersect at $B=0$ in 
three-dimensional space spanned by the spherical coordinates 
$(B,\Theta ,\Phi )$. If the magnetic field traces out 
a curve $C$ starting at ${\bf e} (0)$ and 
ending at ${\bf e} (t)$, the noncyclic geometric 
phase for the eigenstates (\ref{eq:complexeigenstates}) 
becomes 
\begin{eqnarray}
\gamma_{\pm} [C] & = &  
\mp \arg \{ \exp (-i(\Phi (t) -\Phi (0))/2 ) \cos (\Theta (t)/2) 
\cos (\Theta (0)/2) \nonumber \\ 
 & & + \exp (i(\Phi (t) -\Phi (0))/2 ) 
\sin (\Theta (t)/2) \sin (\Theta (0)/2) \} 
\mp \frac{1}{2} \int_{0}^{t}
dt' \, \dot{\Phi} (t') \cos \Theta (t') 
\nonumber \\ 
 & \equiv & \mp 
f ({\bf e} (0),{\bf e} (t)) \mp  
\frac{1}{2} \int_{0}^{t} dt' \, \dot{\Phi} (t') 
\cos \Theta (t') \equiv \pm \gamma [C] .  
\label{eq:compvec}
\end{eqnarray}
In particular it is straightforward to check that for $\Theta (t) 
= \Theta (0) = \pi /2$, (\ref{eq:compvec}) reduces, up to an integer
multiple of $2\pi$, to the geometric phase discussed for the sign 
change problem in Sec. {\bf III}.

Let us discuss how the geometric phase $\gamma [C]$ can 
be measured in a spin-polarization experiment. The idea is 
to analyze the polarization vector 
${\bf P}= \langle \Psi |{\underline{\hat{\sigma}}}|\Psi \rangle$ 
of a neutron beam in the pure state 
$|\Psi \rangle \langle \Psi |$, which has acquired a noncyclic 
geometric phase. Let us concentrate on the $z-$component 
$P_{z}$ of ${\bf P}$. To observe $P_{z}$ one could for example 
split the beam using a Stern-Gerlach field in the $z-$direction. 
Suppose $|\Psi \rangle = c_{+} |+;{\bf e}_{z}\rangle + 
c_{-} |-;{\bf e}_{z}\rangle$ and using that $I_{\pm} \propto 
|c_{\pm}|^{2}$, where $I_{\pm}$ are the intensities of the two 
emergent sub-beams, we then obtain the normalized experimental 
$z-$polarization 
\begin{equation}
P_{z}^{exp} = \frac{1-I_{-}/I_{+}}{1+I_{-}/I_{+}} .  
\end{equation}

We now derive the corresponding theoretical expression for
$P_{z}$ in the geometric phase experiment. 
Suppose the neutrons are prepared in the pure state 
$|+;{\bf e}_{z} \rangle \langle + ;{\bf e}_{z}|$, so  
that the initial polarization in the $z-$direction 
$P_{z} (0)$ equals unity. Assume ${\bf e}$ is taken,  
during the time interval $[0,t ]$, from 
${\bf e} (0) = (\sin \Theta (0) ,0 ,\cos \Theta (0))$ to 
${\bf e} (t ) = (\sin \Theta (t ) \cos \Phi (t ),
\sin \Theta (t )\sin \Phi (t ),\cos \Theta (t ))$. 
Expressing $|\Psi ;0 \rangle = |+;{\bf e}_{z} \rangle$ in 
terms of $|\pm ;{\bf e} (0) \rangle$ using 
(\ref{eq:complexeigenstates}) yields 
\begin{equation}
|\Psi ;0 \rangle =  
\cos ( \Theta (0)/2) |+;{\bf e} (0) \rangle -
\sin (\Theta (0)/2) |-;{\bf e} (0) \rangle ,   
\end{equation} 
and it follows that in the adiabatic approximation the final 
state vector is 
\begin{eqnarray}
|\Psi ;t \rangle & = & 
\exp \left( i\omega_{B}t + \alpha_{+} (t ) \right) 
\cos ( \Theta (0)/2) |+;{\bf e} (t ) \rangle 
\nonumber \\ 
 & & - \exp \left( -i\omega_{B}t + \alpha_{-} (t ) \right) 
\sin (\Theta (0)/2) |-;{\bf e} (t ) \rangle ,   
\label{eq:finalvector}
\end{eqnarray}
where $\alpha_{\pm} (t)$ are given by (\ref{eq:dynamicalphase}) 
for the state vectors $|\pm ;{\bf e} (t) \rangle$. From 
(\ref{eq:finalvector}) and the relations 
\begin{eqnarray}
\langle \, \pm \, ;{\bf e} (t ) |\hat{\sigma}_{z} | 
\pm ;{\bf e} (t ) \rangle & = & 
\pm \cos \Theta (t ) 
\nonumber \\ 
\langle \, \pm \, ;{\bf e} (t ) |\hat{\sigma}_{z} | 
\mp ;{\bf e} (t ) \rangle & = & 
- \sin \Theta (t )  
\end{eqnarray}
we obtain 
\begin{equation}
P_{z} (t ) = \cos \Theta (0) \cos \Theta (t ) - 
\sin \Theta (0) \sin \Theta (t ) 
\cos (2f ({\bf e} (0),{\bf e} (t )) + 
2\omega_{B} t + \gamma_{+} [C] - \gamma_{-} [C]) ,  
\end{equation}
where $\gamma_{+} [C] - \gamma_{-} [C] = 2\gamma [C]$ 
according to (\ref{eq:compvec}). So keeping 
$B,t ,{\bf e} (0)$ and ${\bf e} (t)$ 
fixed for a given initial pure state and varying the 
shape of $C$, the adiabatic noncyclic geometric phase 
alone causes $P_{z} (t )$ to vary. Thus by observing 
$P_{z}$ for different $C$'s, $\gamma [C]$ can be verified 
up to an additive integer multiple of $\pi$. By repeating 
the experiment for different times $t$ but for the same path 
$C$ it is furthermore possible to check the indifference 
of $\gamma [C]$ on how the path $C$ is traversed. Similarly, 
the global phase invariance of the adiabatic noncyclic 
geometric phase can be verified by checking that $\gamma [C]$ 
is independent of $B$.

\section{Conclusions}
We have shown that cyclicity is not a necessary condition 
for the existence of a geometric phase in adiabatic evolution. 
A noncyclic phase of geometric origin has to be included in 
the approximate adiabatic wave function. This phase has been 
shown to be independent of how the path is traversed in 
parameter space and of the phase of the instantaneous energy 
eigenstate $|n;{\bf R}(t)\rangle$. This adiabatic noncyclic 
geometric phase has been frequently overlooked in the literature 
and reasons for this have been analyzed here. The adiabatic 
noncyclic geometric phase has been shown to be nonzero by 
explicit calculations for the sign-change problem and a 
noncyclic Aharonov-Bohm set-up. Moreover these two examples 
show that the noncyclic geometric phase gives a deeper physical 
insight than considering only the cyclic times. We have also 
discussed a spin$-\frac{1}{2}$ experiment to measure the 
adiabatic noncyclic geometric phase and that can be used to 
check its properties.

\section*{Acknowledgements}
We thank Jeeva Anandan, Guido Bacciagaluppi, Harvey R. Brown, 
Henrik Carlsen, Mark S. Child, Osvaldo Goscinski and Sandu 
Popescu for most valuable discussions. G.G. acknowledges an 
EC `Marie Curie' fellowship. E.S. acknowledges financial support 
from The Wenner-Gren Foundation.

\newpage

\setcounter{equation}{0}
\renewcommand{\theequation}{A.\arabic{equation}}
\section*{Appendix}
In this Appendix we compare the approach taken in this paper 
and an approach based on the geodesic closure of the path. 
Although both methods are numerically equivalent the latter 
is shown to have two disadvantages: (a) calculations of the 
geometric phase are much easier in the approach of this 
paper, and most importantly for the adiabatic theorem, (b) 
the geodesic closure will in general lie outside the adiabatic 
manifold. 

It was demonstrated by Samuel and Bhandari \cite{samuel} that 
for a nonadiabatic noncyclic evolution the geometric phase can 
be written as
\begin{equation}
\gamma_{SB} = \oint_{{\cal C}+{\cal C}_{g-c}
+{\cal C}_{v}} A = \int_{S_{g-c}} dA ,  
\label{eq:sbformula}
\end{equation}
where, as shown in Fig. 3, ${\cal C}$ is the open path in Hilbert 
space ${\cal H}$, ${\cal C}_{g-c}$ the horizontal lift of the shortest 
geodesic joining the initial and final points of the projection of ${\cal C}$ 
on ray space ${\cal P}$, i.e. the space where Hilbert space vectors 
differing only in phase are identified, and ${\cal C}_{v}$ the vertical 
path in the phase direction. The quantity $A$ is given by
\begin{equation}
A = \text{Im} \langle \phi | d\phi \rangle 
\end{equation}
with 
\begin{equation}
|\phi ;t \rangle = | \Psi ;t \rangle \exp \left( \frac{i}{\hbar}
\int_{0}^{t} dt' \langle \Psi ;t' |\hat{H} (t')|\Psi ;t' \rangle  
\right)
\end{equation}
and $|\Psi ;t \rangle$ is the Schr\"{o}dinger solution for the 
Hamiltonian operator $\hat{H} (t)$. $S_{g-c}$ is the surface inside 
the projection on ${\cal P}$ of the closed Hilbert space curve 
${\cal C} + {\cal C}_{g-c} + {\cal C}_{v}$, and we have used Stokes' 
theorem for the last equality in (\ref{eq:sbformula}). Moreover 
these authors demonstrated that this phase is equal to the 
Pancharatnam phase $\beta$ \cite{pancharatnam}, i.e.   
\begin{equation}
\gamma_{SB} = \beta = 
\arg \langle \phi ;0 | \phi ;t \rangle . 
\label{eq:pancharatnam}
\end{equation}
Using this last equality it is straightforward to show that 
expression (\ref{eq:generalgp}) for the noncyclic adiabatic phase 
is numerically equal to the Samuel-Bhandari phase 
(\ref{eq:sbformula}) for adiabatic evolution. That is, by choosing 
the global phase for the adiabatic eigenvector 
$|n;{\bf R} (t) \rangle$ such that 
\begin{equation}
\text{Im} \langle n;{\bf R} (t) |\nabla_{{\bf R}}| 
n;{\bf R} (t) \rangle = 0 ,  
\end{equation}
then (\ref{eq:generalgp}) reduces to (\ref{eq:pancharatnam}). 

However the use of the geodesic approach has clear disadvantages. 
On the practical side the evaluation of $\gamma_{SB}$ in 
(\ref{eq:sbformula}) needs the extra calculation of the geodesic. 
More fundamental for the adiabatic theorem is the fact that in 
general the geodesic lies outside the adiabatic manifold. In this 
general case the surface $S_{g-c}$ in (\ref{eq:sbformula}) cannot 
be represented in the space of slow parameters ${\bf R}$. In the 
most favorable case, when the slow parameter space covers the whole 
ray space, Stokes' theorem can be used directly on parameter space. 
An example of this is of two-level systems discussed in Sec. {\bf V}, 
for which also holds that the parameter space metric and the ray space metric 
are the same \cite{berry3}. Therefore Berry's result for cyclic 
adiabatic evolution in two-level systems, 
$\gamma_{\pm} [C_{0}] = \mp \Omega /2$, with $\Omega$ the solid angle 
enclosed by the closed path $C_{0}$, generalizes in the noncyclic 
case to $\gamma_{\pm} [C] = \mp \Omega_{g-c} /2$, with $\Omega_{g-c}$ 
the solid angle defined by the closed path $C + C_{g-c}$, where 
$C_{g-c}$ is the shortest geodesic in parameter space connecting
the end-points of $C$. 

In the following we illustrate this result for two particular examples. 
First consider the case when $\Theta = \pi /2$, i.e. the sign change
case, for which (\ref{eq:compvec}) becomes 
\begin{eqnarray}
\gamma_{\pm} [C] & = & \mp \arg \cos ((\Phi (t)-\Phi (0))/2) \nonumber \\  
 & = & \mp \left\{ \begin{array}{ll} 
0 & {\mbox{if}} \, \, 
\Phi (0)\leq \Phi (t) < \Phi (0)+\pi \\
{\mbox{undefined}} & {\mbox{if}} \, \, 
\Phi (t) = \Phi (0)+\pi \\ 
\pi & {\mbox{if}} \, \, 
\Phi (0)+\pi < \Phi (t) \leq \Phi (0)+2\pi 
\end{array} \right\} =   
\mp \frac{1}{2} \Omega_{g-c} 
\end{eqnarray}
as shown in Fig. 4. Note that when $\Phi (t) = \Phi (0) + \pi$ there 
are infinitely many geodesics connecting the end-points  
and all of them give different values of the geometric
phase which is then undefined. Another interesting example is given 
in Fig. 5 for which (\ref{eq:compvec}) can be checked to give the 
solid angle result 
\begin{equation}
\gamma_{\pm} [C] = \mp \frac{\pi}{4} = \mp \frac{1}{2} \Omega_{g-c} . 
\end{equation}

\newpage
\section*{Figure captions}
Fig. 1. Two electronic potential energy surfaces $E_{\pm} (R)$ 
which conically intersect at the origin $R=0$ in nuclear 
configuration space. 
\vskip 0.5 cm 
Fig. 2. A charged particle confined in an impenetrable 
box by the potential $V$ and transported along $C$ around 
a magnetic flux line. The box is situated at ${\bf R}$ 
and the particle is described by the coordinate ${\bf r}$. 
An asymmetric shape has been chosen and the size of the box 
is such that the overlap between the initial wave function
and the wave function at any later time $t$ is nonvanishing.
The box at the initial time is plotted with thin line.  
We have divided the transport of the box into three 
$\Theta -$intervals: (a) $0 \leq \Theta (t) \leq 2\pi - 
\Delta \theta$, (b) $2\pi - \Delta \theta \leq \Theta (t) 
\leq \Delta \theta$ and (c) $\Theta (t) \geq \Delta \theta$, 
with $\Delta \theta$ the angular length of the boxes. 
\vskip 0.5 cm 
Fig. 3. Illustration of the geodesic closure approach. 
${\cal C}$ is the open path in Hilbert 
space ${\cal H}$, ${\cal C}_{g-c}$ the horizontal lift of the 
shortest geodesic joining the initial and final points of the 
projection of ${\cal C}$ 
on ray space ${\cal P}$, ${\cal C}_{v}$ the vertical path
in the phase direction, and $S_{g-c}$ the surface inside the
projection on ${\cal P}$ of the closed Hilbert space curve 
${\cal C} + {\cal C}_{g-c} + {\cal C}_{v}$.  
\vskip 0.5 cm 
Fig. 4. Noncyclic sign change from the geodesic closure perspective.  
The noncyclic geometric phase is equal to the solid angle defined 
by the path $C$ and its shortest geodesic join (indicated by dashed
line). There are three cases: 
(a) $\Phi (0) \leq \Phi (t) < \Phi (0) + \pi$ with $\gamma_{\pm} [C] = 
\mp \Omega_{g-c} /2 = 0$,
(b) $\Phi (t) = \Phi (0) + \pi$ which has an infinite number of 
geodesic closures making $\gamma_{\pm} [C]$ undefined, and 
(c) $\Phi (0) + \pi < \Phi (t) \leq \Phi (0) + 2\pi$ with 
$\gamma_{\pm} [C] = \mp \Omega_{g-c} /2 = \mp \pi$.  
\vskip 0.5 cm 
Fig. 5. Case for which an open path together with its shortest 
geodesic closure encloses one fourth of the upper hemisphere. 
The adiabatic noncyclic geometric phase is $\gamma_{\pm} = 
\mp \Omega_{g-c} /2 = \mp \pi /4$.  
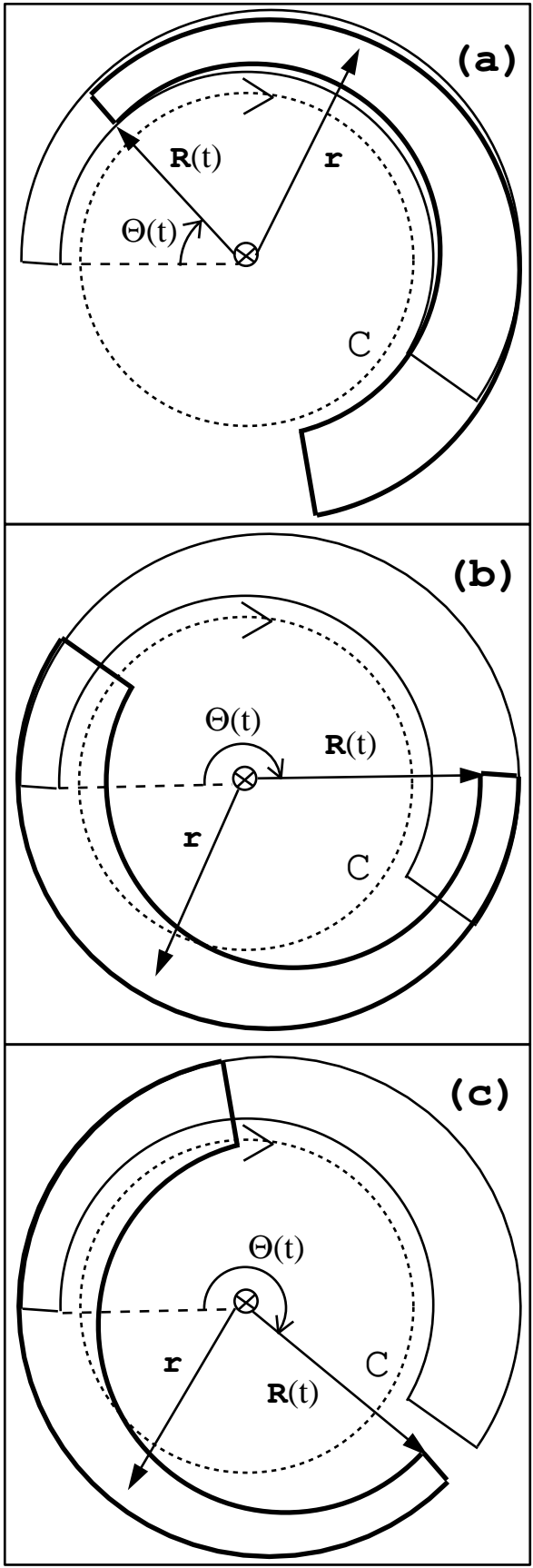
\end{document}